\documentclass[prl,showpacs,twocolumn]{revtex4}
\usepackage{amsmath,graphicx}
\newcount\minute
\newcount\hour
\newcount\hourMins
\def\now
{
  \minute=\time
  \hour=\time \divide \hour by 60
  \hourMins=\hour \multiply\hourMins by 60
  \advance\minute by -\hourMins
  \zeroPadTwo{\the\hour}:\zeroPadTwo{\the\minute}
}
\def\timestamp
{
  \today\ \now
}
\def\today
{
  \zeroPadTwo{\the\day}-\zeroPadTwo{\the\month}-\the\year%
}
\def\zeroPadTwo#1%
{
  \ifnum #1<10 0\fi
  #1%
}
\def \dif {\mathrm{d}}

\pacs{71.10.Pm,73.63.Nm,85.75.-d}
\date{\timestamp}

\begin{document}

\title{Spin-filtering by field dependent resonant tunneling}

\author{Zoran Ristivojevic$^{1,2}$, George I. Japaridze$^{3,4}$ and Thomas Nattermann$^1$}

\affiliation{$^1$Institut f\"{u}r Theoretische Physik, Universit\"{a}t zu K\"{o}ln, Z\"{u}lpicher Stra{\ss}e 77, 50937 K\"{o}ln, Germany}
\affiliation{$^2$Materials Science Division, Argonne National Laboratory, Argonne, IL 60439, USA}
\affiliation{$^3$Andronikashvili Institute of Physics, Tamarashvili 6, 0177 Tbilisi, Georgia}

\affiliation{$^{4}$  Ilia State University, Cholokashvili Ave 3-5, 0162, Tbilisi, Georgia}

\begin{abstract}
We consider theoretically transport in a spinfull
one-channel interacting quantum wire placed in an
external magnetic field. For the case of two
point-like impurities embedded in the wire, under
a small voltage bias the spin-polarized current
occurs at special points in the parameter space,
tunable by a single parameter. At sufficiently
low temperatures complete spin-polarization may
be achieved, provided repulsive interaction between electrons is
not too strong.
\end{abstract}
\maketitle

\textit{Introduction.---} Control and
manipulation of spin degrees of freedom in
nanoscale electronic devices is an active
new field of research
\cite{Prinz95,Zutic+04}. In quantum wires
spin selective transmission of electrons
was considered in the past in a number of
publications
\cite{Schmeltzer+03,Hikihara+05,Kimura+96,Schmeltzer02,Kamide+06}.
In \cite{Schmeltzer+03} a strong asymmetry
of the spin dependent conductances in a
Luttinger liquid (LL) with a magnetic
impurity was observed, which is related to the
Zeeman energy splitting $\Delta$ of the
impurity states.
In \cite{Hikihara+05} the authors consider the
spin dependent backscattering of repulsive
electrons from a single weak impurity in the
presence of a strong magnetic field
$\Delta>\Delta_C\approx 0.2E_F$ where $E_F$
denotes the Fermi energy. Contrary to weak fields,
the backscattering of electrons having spin
parallel to the field may be suppressed making
the impurity transparent, whereas electrons
antiparallel to the field are still reflected.

In the  present paper we report on spin selective
transmission of electrons in a quantum wire
through a quantum dot formed by two impurities.
The mechanism consists in lifting the degeneracy
of the condition for resonant tunneling of up and
down electrons through the quantum dot
\cite{Kane+92b,Furusaki+93b}  by an external magnetic field $H$. Whereas the transmission for
the spin direction which fulfills the resonance
condition is finite for repulsive interaction, it vanishes for the other
spin direction due to the Coulomb blockade in the
quantum dot. The mechanism requires sufficiently
low temperatures such that the Zeeman splitting
$\Delta=g\mu_BH$ and the Coulomb energy of the
quantum dot $\approx E_F/n$ are large compared
to $T$. Here $n $ denotes the number of
electrons in the quantum dot.

\begin{figure}
\includegraphics[width=1\columnwidth]{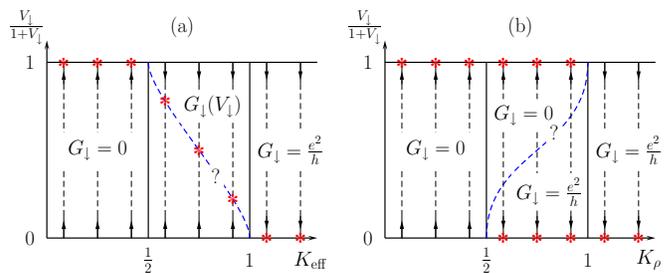}
\caption{The renormalization group flow diagram for $V_\downarrow$
as a function of interaction for parameters when the resonance is
achieved for strong but not for weak impurities (a) and for weak
but not for strong impurities (b). The middle region contains a
line of fixed points in case (a), and a phase transition line in
case (b), precise form of which is unknown. The non-interacting
point is $K_\rho=K_\sigma=K_{\mathrm{eff}}=1$.}\label{Figflow}
\end{figure}

For weak impurities we find a resonance in the region of repulsive
electron interaction where the transmission for one spin direction
is perfect, provided the impurity is weaker than a critical value,
whereas the other spin direction is completely blocked. For strong
impurities transmission is found to change smoothly from perfect
to zero when the interaction strength is increased. As a
difference to the case $H=0$ considered in \cite{Kane+92b}, we
find that the resonance condition for $H\neq0$ is not same in the
two limiting cases of strong and weak impurities and leads to two
scenarios shown in Fig.~\ref{Figflow}.

A similar setup, but under very different
conditions, has been considered recently in
\cite{Kamide+06}. There, the Coulomb blockade
effect was ignored and the magnetic field was assumed to be
unrealistically strong, $\Delta=\mathcal{O}(E_F)$.


\textit{Model.---} We consider electrons in a one dimensional
wire along $x$ axis exposed to an external magnetic field.
Since electrons are confined in the directions transverse to $x$,
orbital effects are suppressed and the only field effect of the
magnetic field is to polarize the  electrons. In the noninteracting case the Zeeman energy splits the Fermi momentum $k_{F,s}$ ($s=\uparrow,\downarrow$)
of the up and down spin electrons by
$|k_{F\uparrow}-k_{F\downarrow}|/(k_{F\uparrow}+k_{F\downarrow})
\approx \Delta/E_F\ll1 $.
The Hamiltonian for electrons in the external impurity potential
$V(x)$ can be described by the Tomonaga-Luttinger
model
\begin{align}\label{H}
&H=\sum_s\int\dif x\bigg\{-i\hbar v_F\left[\psi_{Rs}^\dagger
\partial_x\psi_{Rs}-\psi_{Ls}^\dagger\partial_x\psi_{Ls}\right]\\\notag
&+V(x)\rho_s(x)\bigg\}+\frac{1}{2}\sum_{s,s'}\int\dif x\dif x'W(x-x')\rho_s(x)\rho_{s'}(x'),
\end{align}
where $\psi_{Rs}(x),\psi_{Ls}(x)$ are the
annihilation operators for right- and left-moving
spin-$s$ electrons, $\psi_s=\psi_{Rs}+\psi_{Ls}$ is
the annihilation operator for spin-$s$ electrons,
$\rho_s=\psi_s^\dagger\psi_s$ is the spin-$s$
electron density, and $W(x-x')$ is the screened
Coulomb interaction between electrons
\cite{footnote1}.


We first consider the system without impurities.
Then the model (\ref{H}) describes an interacting
quantum wire with four Fermi points \cite{Gogolin+}. In that
situation it is useful to split terms arising
from the interaction  into inter-subband and
intra-subband terms \cite{Starykh+00}. For repulsive and spin independent interaction
electrons stay in their bands during scattering
processes and the only allowed intersubband
process is the forward scattering
\cite{Gangadharaiah+08}.
While mutually noninteracting subsystems
consisting of spin up and spin down electrons
are described in the bosonized representation by
the standard LL Euclidean action\cite{Giamarchi}
in terms of bosonic fields
$\varphi_\uparrow,\varphi_\downarrow$  with the
Luttinger parameter (LP)
$K=\left(1+\frac{\widetilde{W}(0)-\widetilde{W}(2k_F)}{\pi\hbar
v_F}\right)^{-1/2}$, the intersubband interaction
is diagonalized in symmetric
$\varphi_{\rho}=(\varphi_\uparrow+\varphi_\downarrow)/\sqrt{2}$
and antisymmetric
$\varphi_{\sigma}=(\varphi_\uparrow-\varphi_\downarrow)/\sqrt{2}$
combinations. $\varphi_\rho$ describe
charge and $\varphi_\sigma$ spin degrees of freedom.
The action of the system in the absence of impurities is then given by (for details see \cite{Ristivojevic09,footnote11})
\begin{align}\label{S0model}
&\frac{S_{0}}{\hbar}\!=\!\!\sum_{\ell=\rho,\sigma}\frac{1}{2\pi
K_\ell}\!\int\dif
x\dif\tau\left[\frac{1}{v_\ell}(\partial_\tau\varphi_{\ell})^2
+v_\ell(\partial_x\varphi_{\ell})^2\right],
\end{align}
where
$K_\ell=K\left(1\pm\frac{K^2\widetilde{W}(0)}{\pi\hbar
v_F}\right)^{-1/2}$ with the convention that the
upper (lower) sign corresponds to $\ell=\rho
(\sigma)$. The velocities of excitations are
$v_{\ell}=v_F/K_{\ell}$, where $v_F$ is the Fermi
velocity.

Non-trivial effects come from impurities. We consider two point-like impurities, modeled as
$\delta$-functions of the strength
$V$ and placed at $\pm a/2$. Introducing the
displacement fields at the impurity positions
$\phi_{1s}(\tau)=\varphi_s(-a/2,\tau)$ and
$\phi_{2s}(\tau)=\varphi_s(a/2,\tau)$, the
bosonized form of electron-impurity interaction
reads \cite{Kane+92b,Ristivojevic09} $ {S_1}{}=
\sum_{s}({Vk_{Fs}}/{\pi})\int\dif\tau\left[
\cos\left(2\phi_s+k_{Fs}a\right)+\cos\left(2\phi_s-k_{Fs}a\right)\right].$

To analyze the full action $S_0+S_1$ it is useful
to integrate out degrees of freedom outside the
impurities. In that way one gets an action in
terms of four fluctuating fields in imaginary
time. For low frequencies, $|\omega|\ll v_\ell/a$,
the effective action reads
\begin{align}\label{Seff}
S_{\mathrm{eff}}=\sum_{\ell=\rho,\sigma}\sum_{k=\pm}
\int\frac{\dif\omega}{16\pi^2}
\frac{\hbar|\omega|}{ K_\ell} \left |\Phi_{\ell
k}(\omega)\right|^2 +\int\dif\tau
V_{\mathrm{eff}},
\end{align}
where effective potential energy $V_{\mathrm{eff}}$ reads
\begin{align}\label{Veff}
&V_{\mathrm{eff}}(\phi_{1\uparrow},\phi_{2\uparrow},\phi_{1\downarrow},
\phi_{2\downarrow})=\sum_{\ell}\frac{1}{2}U_\ell\Phi_{\ell^-}(\tau)^2\\\notag&+\sum_{s}V_{s}\left[\cos(2\phi_{1s}+k_{Fs}a)
+\cos(2\phi_{2s}-k_{Fs}a)\right].
\end{align}
Here we have introduced $U_\ell=\frac{\hbar
v_\ell}{2\pi aK_\ell}$, $V_s={Vk_{Fs}}/{\pi}$ and
the fields $\Phi_{\ell k}=\phi_{2\uparrow}+
k\phi_{1\uparrow}\pm(\phi_{2\downarrow}+k\phi_{1\downarrow})$
where $k=\pm$ and our sign convention for
$\ell=\rho,\sigma$ applies.
$\Phi_{\rho^-}$ and $\Phi_{\sigma^-}$ determine
the total charge and spin, respectively, between
the impurities.


The effective potential energy (\ref{Veff})
consists of two types of terms: the charging
energy $E_C=\sum_\ell U_\ell\Phi_{\ell^-}^2/2$
suppresses the accumulation of charge and spin on
the island between impurities, while the
$V_s$-term tends to pin the displacement fields
at the impurity positions. The part
$|\omega||\Phi_{\ell^-}|^2$ of the action
(\ref{Seff}) is a  fluctuation correction to
$E_C$ and is important at resonance points for
strong impurities, when $\Phi_{\ell^-}$ are
undetermined, see below.

In the following we will examine the system
described by (\ref{Seff}) in two limiting cases,
for strong and weak impurity strengths.
In the realistic case of repulsive interaction, we have
$K_\rho<1, K_\sigma>1$ and $U_\sigma<U_\rho$. We
study the model at zero temperature, while
influence of temperature is briefly considered at
the end. Our strategy is to first determine the
ground state from
$V_{\mathrm{eff}}$ without fluctuations, see Eq.~(\ref{Seff}) and then
to include fluctuations in order to check the stability of
that ground state.

\begin{figure}
\includegraphics[width=0.9\columnwidth]{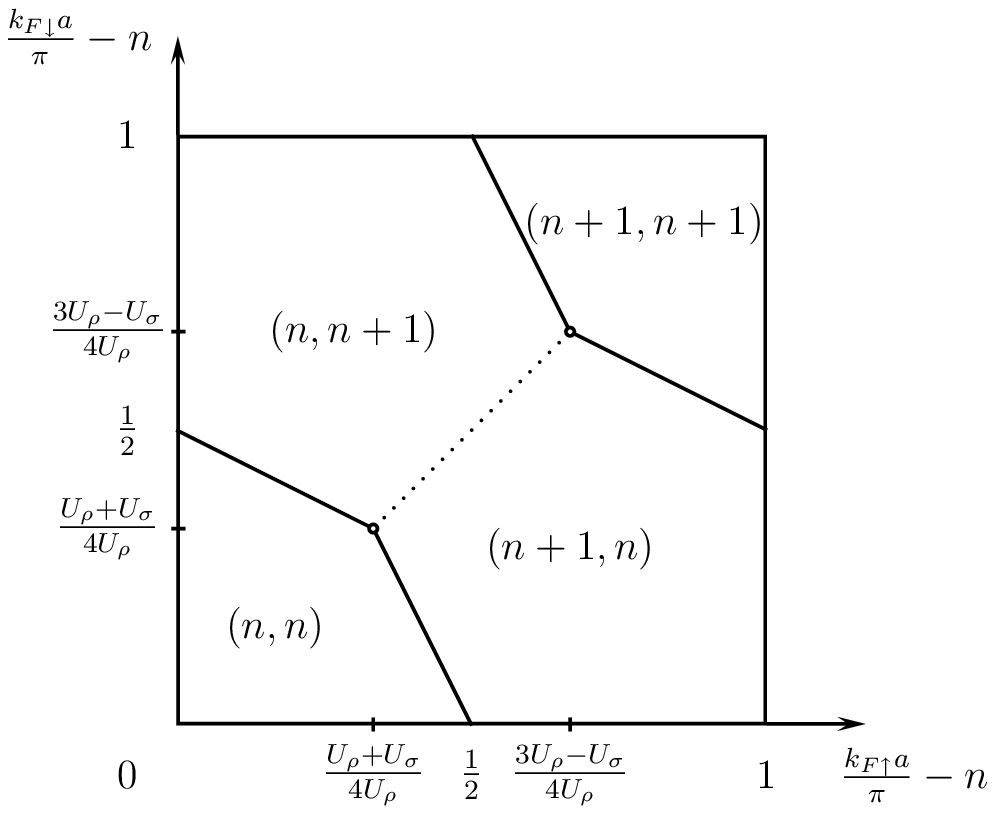}
\caption{Ground state energy configurations of
the charging $E_C(n_\uparrow,n_\downarrow)$ for
repulsive interaction $U_\sigma<U_\rho$. Points
at boundaries between different ground state
configurations correspond to the resonance
points, special points where the ground state
degeneracy is present. Boundaries with solid
lines describe resonances for either up or down
spins, while the dotted line is the
Kondo-resonance for spin exchange process at the
island which exists without magnetic
field, when $k_{F\uparrow}=k_{F\downarrow}$.}\label{Fig}
\end{figure}

\textit{Strong impurities.---}In the limit of
very strong barriers, $V_\uparrow,V_\downarrow\gg
U_\rho,U_\sigma,E_F$, the ground state of the
system is defined by subsequent minimization of
the pinning  and the charging energy, see
Eq.~(\ref{Veff}). The pinning energy terms are
minimal for $2\phi_{ps}=(-1)^p
k_{Fs}a+\pi(1-2n_{ps}),\ p=1,2 $ where $n_{ps}$
are integers. The high degeneracy of the  pinning
energy   is broken by the charging energy.
Plugging $\phi_{ps}$ into $E_C$ and defining
$n_s=n_{2s}-n_{1s}$ one gets
\begin{align}\label{Ec}
E_C(n_\uparrow,n_\downarrow)=&\frac{U_\rho}{2}\left({k_{F\uparrow}a}
+{k_{F\downarrow}a}-{\pi}(n_\uparrow+n_\downarrow)\right)^2\\\notag
&+\frac{U_\sigma}{2}\left({k_{F\uparrow}a}
-{k_{F\downarrow}a}-{\pi}(n_\uparrow-n_\downarrow)\right)^2.
\end{align}
To characterize different nonequivalent minima of (\ref{Ec}) it is
useful to restrict the Fermi momenta to satisfy
$n<k_{F\uparrow}a/\pi,k_{F\downarrow}a/\pi\le n+1$, where $n\ge 0$
is an integer. This implies $n\le n_\uparrow,n_\downarrow\le n+1$.
The  particle number on the island is
$n_{\uparrow}+n_{\downarrow}$. The ground states resulting from
the minimization of the charging energy (\ref{Ec}) are shown in
Fig.~\ref{Fig}. For generic values of $k_{Fs}a$, the ground state
is uniquely determined. However, at special lines different ground
states meet. These lines define the {\it resonance conditions}:
while the number of particles on the island with one spin
direction is fixed at the same value on both sides of the
boundary, the number of electrons with the opposite spins changes
by $\pm 1$. $E_C(n_{\uparrow},n_{\downarrow})=E_C(n_{\uparrow}\pm
1,n_{\downarrow})$ and
$E_C(n_{\uparrow},n_{\downarrow})=E_C(n_{\uparrow},n_{\downarrow}\pm
1)$ are the resonance conditions for the up and down spin
electrons, respectively. As a result a particle having the
degenerate spin can tunnel through the quantum dot in a sequential
tunneling process without changing its energy.
Hence we have a spin-selective barrier
transparency.

We further solve the model along the boundary line where $E_C(n+1,n)=E_C(n+1,n+1)$. Similar results hold for other cases. The fields $\phi_{p\uparrow},\ p=1,2$ are locked by the strong impurity pinning and have fixed values of $n_\uparrow$. Approximating the nonlinear cosine term by a quadratic term for the  $\phi_{p\uparrow}$ one can integrate out them from the action (\ref{Seff})\cite{footnote2}.
The resulting effective action then reads
\begin{align}\label{Seff-boundary}
{S_{\mathrm{eff}}'}=&\int\frac{\dif\omega}{4\pi^2}\frac{\hbar|\omega|}{K_{\mathrm{eff}}}
\left(\left|\phi_{1\downarrow}(\omega)|^2+|\phi_{2\downarrow}(\omega)\right|^2\right)
\\\notag&+\int\dif\tau
V_{\mathrm{eff}}
\left(-\frac{k_{F\uparrow}a+\pi}{2},
\frac{k_{F\uparrow}a+\pi}{2},\phi_{1\downarrow},\phi_{2\downarrow}\right),
\end{align}
with $K_{\mathrm{eff}}=\frac{2K_\rho
K_\sigma}{K_\rho+K_\sigma}$.
It describes the resonant tunneling of spin down electrons 
and is analogous to the case of spinless electrons
\cite{Kane+92b}. The partition function is dominated by tunneling
events connecting degenerate minima of the strong impurity
potential. Using the Coulomb gas representation
\cite{Kane+92a,Kane+92b,Furusaki+93a} one can produce the
renormalization group equations for the tunneling transparency
$t_\downarrow$ of barriers for spin-$\downarrow$ electrons. For
strong impurity potential $V_\downarrow$ it reads $d_l
t_\downarrow=t_\downarrow\left(1-1/(2K_{\mathrm{eff}})\right)$,
from which we get that for $K_{\mathrm{eff}}>\frac{1}{2}$ the
transparency $t_\downarrow$ increases, or equivalently, the
strength of $V_\downarrow$ flows to smaller values at low
energies. Outside the resonance lines, $t_s$ flows to zero for any
repulsion, similar to the single impurity case.

\textit{Weak impurities.---}In the limit of weak impurities, $V_\uparrow,V_\downarrow\ll U_\rho,U_\sigma$, the action (\ref{Seff}) is minimized for $\Phi_{\ell^-}=0$.
This corresponds to fixed charge and spin on the island. Integrating out the   $\Phi_{\ell^-}$ fluctuations
from (\ref{Seff}), new scattering processes of the form
$\sum_{s}2V_s\cos(k_{Fs}a)\cos(\phi_{1s}+\phi_{2s})+V^{(2)}\sin(k_{F\uparrow}a)\sin(k_{F\downarrow}a) \cos\Phi_{\rho^+}$ are generated, where $V^{(2)}=V_\uparrow
V_\downarrow \frac{U_\sigma-U_\rho}{2U_\rho
U_\sigma}$. Other generated higher order
processes are irrelevant for repulsive
interaction. The resonance condition for the
spin-$s$ particles is now given by $\cos
(k_{Fs}a)=0$.

For the generic situation $\cos(k_{Fs}a)\neq 0$,
the single electron backscattering processes are
the most important ones. To leading order in the impurity potential, the
renormalization group (RG) flow equations is
$\dif_l
V_s=V_s\left[1-(K_{\rho}+K_\sigma)/2\right],$
from which we conclude that backward scattering
terms $V_s$ are relevant for
$K_\rho+K_\sigma<2$.
Since the point impurity is a local quantity it
can not renormalize bulk quantities such as
$K_\rho,K_\sigma$, and the flow of $V_s$ is
vertical \cite{Fisher+85,Kane+92b}. The flow
diagram for $V_\downarrow$ is shown in
Fig.~\ref{Figflow}a. Since the two limiting cases
have opposite flow, it is plausible to expect a
line of attractive fixed points somewhere in
between, corresponding to a new phase, where
spins of one direction (here down spins) have
nonzero transmission at zero temperature, while
the other spin direction is blocked \cite{footnote3}.

In the resonance case $\cos(k_{F\downarrow}a)=0$ and
$|\cos(k_{F\uparrow}a)|\notin \{0,1\}$, the
two-particle scattering processes should be taken into
account (only for spin-$\downarrow$ electrons since
spin-$\uparrow$ already have backscattering in the lowest nonvanishing order).
From the RG flow equation $\dif_l
V^{(2)}=V^{(2)}(1-2K_\rho),$ we conclude that
spin-$\downarrow$ electrons are effectively free at
low energies for $K_\rho>1/2$. In Fig.~\ref{Figflow}b we show the flow of
$V_\downarrow$. Again the flows of
two limiting cases are opposite, resulting in a
separatrix in between the two resulting phases:
perfectly conducting for spin down for small
enough $V_\downarrow$ and insulating for larger
$V_\downarrow$. Spin up electrons are always in the insulating phase in that case. Outside the middle region, the
flow of $V_\downarrow$ is as in the single
impurity case: toward zero for attractive
interaction and toward infinity for very
repulsive interaction.


In order to check the correctness of the assumption of massive
fluctuations for the $\phi_{p\uparrow}$ fields made when we
derived (\ref{Seff-boundary}), we will  examine (\ref{Seff}) in
cases when the $\phi_{p\downarrow}$ fields are either freely
fluctuating or completely frozen, limits that are appropriate
close to the non-interacting point and in the strongly repulsive
region, respectively, see Fig.~\ref{Figflow}a. Integrating out
$\phi_{p\downarrow}$ one gets an action that matches the action of
a single impurity in LL. In the former case with the LP
$(K_\rho+K_\sigma)/2$, and in the latter with $K_{\mathrm{eff}}$.
In both cases any repulsion ultimately renormalizes $V_\uparrow$
to infinity, which justifies massive fluctuations of
$\phi_{p\uparrow}$.


\textit{Transport.---} Now we will consider the conductance of our
system using the anticipated flow diagram, see Fig.~\ref{Figflow}.
Assuming the applied voltage across the dot is $V_G$ and at the
ends of the wire is $V_L$, an additional term should be included
in the action (\ref{Seff}), which reads
$-eV_G\int\dif\tau\Phi_{\rho^-}/\pi-
eV_L\int\dif\tau\Phi_{\rho^+}/(2\pi)$. The voltage $V_L$ pushes
the electrons to advance in one direction along the wire, while
the gate voltage $V_G$ serves as a single tuning parameter. Due to
nonzero $V_G$, the shifted Fermi momenta
$k_{Fs}'=k_{Fs}-\frac{eV_GK_\rho^2}{\hbar v_F}$ should be taken in
the above results, e.g., for the resonance conditions. This means
the latter can be achieved by adjusting $V_G$ for fixed both
magnetic field and distance between impurities.

Without impurities or for attractive interaction in the low energy
limit the system is described by Eq.~(\ref{S0model}) and has the
perfect non-spin-polarized conductance
$G_\uparrow=G_\downarrow=e^2/h$
\cite{Maslov+95,Ponomarenko95,Safi+95}. The situation drastically
changes when impurities are present. In the non-resonant case, our
model translates into the single impurity problem with the LP
$K_{\mathrm{eff}}$. Therefore, the conductance is suppressed at
low $V_L$ for repulsive interaction for both spin directions as
$\sim V_L^{2/K_{\mathrm{eff}}-2}$.

On the resonance that corresponds to Fig.~\ref{Figflow}a, i.e.~
for strong impurities when the charge state for spin-$\downarrow$
electrons is degenerate on the island, one gets spin-polarized
conductance. Inside the region where the new line of fixed points
appears, different scattering is experienced by two spin
orientations. While $G_\uparrow$ is suppressed at low voltages as
$\sim V_L^{2/K_{\mathrm{eff}}-2}$ near the point
$K_{\mathrm{eff}}=1/2$, and as $\sim V_L^{4/(K_\rho+K_\sigma)-2}$
for $K_\rho+K_\sigma\to 2^-$, $G_\downarrow$ is not suppressed
even at very low voltages. It is controlled by the fixed point
value $V_\downarrow^*(K_{\mathrm{eff}})$ which determines the
effective strength of impurity scattering for a given
$K_{\mathrm{eff}}$. We can estimate the conductance as
$G_\downarrow(K_{\mathrm{eff}})
\approx\frac{e^2}{h}\frac{1}{1+[\pi
V_\downarrow^*(K_{\mathrm{eff}})/E_F]^2}$. Within our approach we
are not able to determine $V_\downarrow^*(K_{\mathrm{eff}})$. We
expect that the fermionic method used in Ref.~\cite{Matveev+93},
which is beyond the scope of the present paper, could give more
results.

On the resonance that corresponds to weak
impurities, Fig.~\ref{Figflow}b, the system again
has spin-polarized conductance which is
controlled by the fixed points. In the lowest
non-trivial order we have $G_\downarrow=e^2/h$
for $K_\rho>1/2$ and $G_\uparrow\sim
V_L^{4/(K_\rho+K_\sigma)-2}$, for not too big
initial values of impurity strengths. Otherwise
the spin polarization is destroyed and the
conductance behaves as in the non-resonant case.

So far we considered zero temperatures. At finite temperature the
picture will be qualitatively unchanged until the electron thermal
energy is much smaller than the charging and Zeeman energy. In the
opposite case, which is the high frequency limit $|\omega|\gg
v_\ell/a$, or $T\gg K_\ell U_\ell$, for the starting action one
would get Eqs.~(\ref{Seff}) and (\ref{Veff}) with the replacements
$K_\ell\to K_\ell/2,U_\ell\to 0$. Then the coherent effects of
impurities are missing and our system effectively has the single
impurity behavior \cite{Kane+92b,Furusaki+93a}.

\textit{Conclusions.---} We have shown that a
quantum wire with two impurities in an external
magnetic field may have spin-filter properties
for repulsive interaction. Our study is based on
the resonance tunneling phenomenon which may be
tuned by a single parameter for only one spin
polarization.

We thank L. Chen and A. Petkovi\'{c} for
helpful discussions. This work is
supported by the Deutsche
Forschungsgemeinschaft under
the grant NA222/5-2 and through SFB 608 (D4).


\begin{thebibliography}{99}
\expandafter\ifx\csname natexlab\endcsname\relax\def\natexlab#1{#1}\fi
\expandafter\ifx\csname bibnamefont\endcsname\relax
  \def\bibnamefont#1{#1}\fi
\expandafter\ifx\csname bibfnamefont\endcsname\relax
  \def\bibfnamefont#1{#1}\fi
\expandafter\ifx\csname citenamefont\endcsname\relax
  \def\citenamefont#1{#1}\fi
\expandafter\ifx\csname url\endcsname\relax
  \def\url#1{\texttt{#1}}\fi
\expandafter\ifx\csname urlprefix\endcsname\relax\def\urlprefix{URL }\fi
\providecommand{\bibinfo}[2]{#2}
\providecommand{\eprint}[2][]{\url{#2}}

\bibitem[{\citenamefont{Prinz}(1995)}]{Prinz95}
\bibinfo{author}{\bibfnamefont{G.~A.} \bibnamefont{Prinz}},
  \bibinfo{journal}{Phys. Today} \textbf{\bibinfo{volume}{48}},
  \bibinfo{pages}{58} (\bibinfo{year}{1995}).

\bibitem[{\citenamefont{\v{Z}uti\'{c} et~al.}(2004)\citenamefont{\v{Z}uti\'{c},
  Fabian, and \mbox{Das Sarma}}}]{Zutic+04}
\bibinfo{author}{\bibfnamefont{I.}~\bibnamefont{\v{Z}uti\'{c}}} \emph{et al.},
\bibinfo{journal}{Rev. Mod. Phys.} \textbf{\bibinfo{volume}{76}},
  \bibinfo{pages}{323} (\bibinfo{year}{2004}).

\bibitem[{\citenamefont{Schmeltzer et~al.}(2003)\citenamefont{Schmeltzer,
  Bishop, Saxena, and Smith}}]{Schmeltzer+03}
\bibinfo{author}{\bibfnamefont{D.}~\bibnamefont{Schmeltzer}} \emph{et al.},
\bibinfo{journal}{Phys. Rev. Lett.} \textbf{\bibinfo{volume}{90}},
  \bibinfo{pages}{116802} (\bibinfo{year}{2003}).

\bibitem[{\citenamefont{Hikihara et~al.}(2005)\citenamefont{Hikihara, Furusaki,
  and Matveev}}]{Hikihara+05}
\bibinfo{author}{\bibfnamefont{T.}~\bibnamefont{Hikihara}} \emph{et al.},
\bibinfo{journal}{Phys. Rev. B} \textbf{\bibinfo{volume}{72}},
  \bibinfo{pages}{035301} (\bibinfo{year}{2005}).

\bibitem[{\citenamefont{Kimura et~al.}(1996)\citenamefont{Kimura, Kuroki, and
  Aoki}}]{Kimura+96}
\bibinfo{author}{\bibfnamefont{T.}~\bibnamefont{Kimura}} \emph{et al.},
\bibinfo{journal}{Phys. Rev. B} \textbf{\bibinfo{volume}{53}},
  \bibinfo{pages}{9572} (\bibinfo{year}{1996}).

\bibitem[{\citenamefont{Kamide et~al.}(2006)\citenamefont{Kamide, Tsukada, and Kurihara}}]{Kamide+06}
\bibinfo{author}{\bibfnamefont{K.}~\bibnamefont{Kamide}} \emph{et al},
\bibinfo{journal}{Phys. Rev. B} \textbf{\bibinfo{volume}{73}},
\bibinfo{pages}{235326} (\bibinfo{year}{2006}).

\bibitem[{\citenamefont{Schmeltzer}(2002)}]{Schmeltzer02}
\bibinfo{author}{\bibfnamefont{D.}~\bibnamefont{Schmeltzer}},
  \bibinfo{journal}{Phys. Rev. B} \textbf{\bibinfo{volume}{65}},
  \bibinfo{pages}{193303} (\bibinfo{year}{2002}).

\bibitem[{\citenamefont{Kane and Fisher}(1992{\natexlab{a}})}]{Kane+92b}
\bibinfo{author}{\bibfnamefont{C.~L.} \bibnamefont{Kane}} \bibnamefont{and}
  \bibinfo{author}{\bibfnamefont{M.~P.~A.} \bibnamefont{Fisher}},
  \bibinfo{journal}{Phys. Rev. B} \textbf{\bibinfo{volume}{46}},
  \bibinfo{pages}{15233} (\bibinfo{year}{1992}{\natexlab{a}}).

\bibitem[{\citenamefont{Furusaki and
  Nagaosa}(1993{\natexlab{a}})}]{Furusaki+93b}
\bibinfo{author}{\bibfnamefont{A.}~\bibnamefont{Furusaki}} \bibnamefont{and}
  \bibinfo{author}{\bibfnamefont{N.}~\bibnamefont{Nagaosa}},
  \bibinfo{journal}{Phys. Rev. B} \textbf{\bibinfo{volume}{47}},
  \bibinfo{pages}{3827} (\bibinfo{year}{1993}{\natexlab{a}}).

\bibitem[{\citenamefont{Chang and Schmeltzer}(2005)}]{Chang+05}
\bibinfo{author}{\bibfnamefont{H.-Y.} \bibnamefont{Chang}} \bibnamefont{and}
  \bibinfo{author}{\bibfnamefont{D.}~\bibnamefont{Schmeltzer}},
  \bibinfo{journal}{Phys. Lett. A} \textbf{\bibinfo{volume}{345}},
  \bibinfo{pages}{45} (\bibinfo{year}{2005}).



\bibitem{footnote1} Here we have assumed the same
Fermi velocity for up and down spins. This is the
case, e.g., in carbon nanotubes. For particles
described by a linearized quadratic dispersion
relation, the relative split of the Fermi velocity
is of the same order as that of the momenta. When this splitting is taken into account, it leads to different scaling dimensions $\gamma_s$ of the single impurity operator for spin-$s$ electrons, with $|\gamma_\uparrow-\gamma_\downarrow|=\mathcal{O}(\Delta/E_F)\ll 1$, which leads to the spin-polarized transmission in a very narrow region near the noninteracting point, see Refs.~\cite{Kimura+96,Schmeltzer02,Chang+05}. The effect of Fermi velocity asymmetry is inessential for our considerations, since we find, even for very small magnetic field, regions of finite size that have spin-polarized transmission, see further text and Fig.~\ref{Figflow}.




\bibitem[{\citenamefont{Gogolin et~al.}(1998)\citenamefont{Gogolin, Nersesyan,
  and Tsvelik}}]{Gogolin+}
\bibinfo{author}{\bibfnamefont{A.}~\bibnamefont{Gogolin}},
  \bibinfo{author}{\bibfnamefont{A.}~\bibnamefont{Nersesyan}},
  \bibnamefont{and} \bibinfo{author}{\bibfnamefont{A.}~\bibnamefont{Tsvelik}},
  \emph{\bibinfo{title}{Bosonization and Strongly Correlated Systems}}
  (\bibinfo{publisher}{Cambridge University Press}, \bibinfo{year}{1998}).

\bibitem[{\citenamefont{Starykh et~al.}(2000)\citenamefont{Starykh, Maslov,
  H\"{a}usler, and Glazman}}]{Starykh+00}
\bibinfo{author}{\bibfnamefont{O.~A.} \bibnamefont{Starykh}} \emph{et al.}, in \emph{\bibinfo{booktitle}{Low-Dimensional
  Systems}}, edited by
  \bibinfo{editor}{\bibfnamefont{T.}~\bibnamefont{Brandeis}}
  (\bibinfo{publisher}{Springer, New York}, \bibinfo{year}{2000}),
  p.~\bibinfo{pages}{37}.

\bibitem[{\citenamefont{Gangadharaiah et~al.}(2008)\citenamefont{Gangadharaiah,
  Sun, and Starykh}}]{Gangadharaiah+08}
\bibinfo{author}{\bibfnamefont{S.}~\bibnamefont{Gangadharaiah}},
  \bibinfo{author}{\bibfnamefont{J.}~\bibnamefont{Sun}}, \bibnamefont{and}
  \bibinfo{author}{\bibfnamefont{O.~A.} \bibnamefont{Starykh}},
  \bibinfo{journal}{Phys. Rev. B} \textbf{\bibinfo{volume}{78}},
  \bibinfo{pages}{054436} (\bibinfo{year}{2008}).

\bibitem[{\citenamefont{Giamarchi}(2003)}]{Giamarchi}
\bibinfo{author}{\bibfnamefont{T.}~\bibnamefont{Giamarchi}},
  \emph{\bibinfo{title}{Quantum Physics in One Dimension}}
  (\bibinfo{publisher}{Clarendon Press, Oxford}, \bibinfo{year}{2003}).

\bibitem[{\citenamefont{Ristivojevic}(2009)}]{Ristivojevic09}
\bibinfo{author}{\bibfnamefont{Z.}~\bibnamefont{Ristivojevic}}, Ph.D. thesis,
  \bibinfo{school}{Cologne University} (\bibinfo{year}{2009}).

\bibitem[{\citenamefont{Shytov et~al.}(2003)\citenamefont{Shytov, Glazman, and Starykh}}]{Shytov+03}
\bibinfo{author}{\bibfnamefont{A.~V.} \bibnamefont{Shytov}} \emph{et al.},
  \bibinfo{journal}{Phys. Rev. Lett.} \textbf{\bibinfo{volume}{91}},
  \bibinfo{pages}{046801} (\bibinfo{year}{2003}).

\bibitem{footnote11} The electron backscattering term is an irrelevant operator for repulsive interaction and does not influence the low energy physics. In principle, its effects may be considered as done in Ref.~\cite{Shytov+03}.


\bibitem{footnote2} The same result is obtained also by
assuming constant $2\phi_{p\uparrow}=\mp(k_{F\uparrow}a+\pi)$ and
eliminating them from Eq.~(\ref{Seff}) using the
identity $\int\dif\omega|\omega||\phi(\omega)|^2=\int\dif\tau\dif\tau'
\frac{[\phi(\tau)-\phi(\tau')]^2}{(\tau-\tau')^2}$.


\bibitem[{\citenamefont{Kane and Fisher}(1992{\natexlab{b}})}]{Kane+92a}
\bibinfo{author}{\bibfnamefont{C.~L.} \bibnamefont{Kane}} \bibnamefont{and}
  \bibinfo{author}{\bibfnamefont{M.~P.~A.} \bibnamefont{Fisher}},
  \bibinfo{journal}{Phys. Rev. B} \textbf{\bibinfo{volume}{46}},
  \bibinfo{pages}{7268} (\bibinfo{year}{1992}{\natexlab{b}}).

\bibitem[{\citenamefont{Furusaki and
  Nagaosa}(1993{\natexlab{b}})}]{Furusaki+93a}
\bibinfo{author}{\bibfnamefont{A.}~\bibnamefont{Furusaki}} \bibnamefont{and}
  \bibinfo{author}{\bibfnamefont{N.}~\bibnamefont{Nagaosa}},
  \bibinfo{journal}{Phys. Rev. B} \textbf{\bibinfo{volume}{47}},
  \bibinfo{pages}{4631} (\bibinfo{year}{1993}{\natexlab{b}}).

\bibitem[{\citenamefont{Fisher and Zwerger}(1985)}]{Fisher+85}
\bibinfo{author}{\bibfnamefont{M.~P.~A.} \bibnamefont{Fisher}}
  \bibnamefont{and} \bibinfo{author}{\bibfnamefont{W.}~\bibnamefont{Zwerger}},
  \bibinfo{journal}{Phys. Rev. B} \textbf{\bibinfo{volume}{32}},
  \bibinfo{pages}{6190} (\bibinfo{year}{1985}).

\bibitem[{\citenamefont{Maslov and Stone}(1995)}]{Maslov+95}
\bibinfo{author}{\bibfnamefont{D.}~\bibnamefont{Maslov}} \bibnamefont{and}
  \bibinfo{author}{\bibfnamefont{M.}~\bibnamefont{Stone}},
  \bibinfo{journal}{Phys. Rev. B} \textbf{\bibinfo{volume}{52}},
  \bibinfo{pages}{R5539} (\bibinfo{year}{1995}).

\bibitem[{\citenamefont{Ponomarenko}(1995)}]{Ponomarenko95}
\bibinfo{author}{\bibfnamefont{V.~V.} \bibnamefont{Ponomarenko}},
  \bibinfo{journal}{Phys. Rev. B} \textbf{\bibinfo{volume}{52}},
  \bibinfo{pages}{R8666} (\bibinfo{year}{1995}).

\bibitem[{\citenamefont{Safi and Schulz}(1995)}]{Safi+95}
\bibinfo{author}{\bibfnamefont{I.}~\bibnamefont{Safi}} \bibnamefont{and}
  \bibinfo{author}{\bibfnamefont{H.~J.} \bibnamefont{Schulz}},
  \bibinfo{journal}{Phys. Rev. B} \textbf{\bibinfo{volume}{52}},
  \bibinfo{pages}{R17040} (\bibinfo{year}{1995}).


\bibitem[{\citenamefont{Yi and Kane}(1996)}]{Yi+96}
\bibinfo{author}{\bibfnamefont{H.}~\bibnamefont{Yi}} \bibnamefont{and}
  \bibinfo{author}{\bibfnamefont{C.~L.} \bibnamefont{Kane}},
  \bibinfo{journal}{arxiv:cond-mat/9602099}  (\bibinfo{year}{1996}).

\bibitem[{\citenamefont{Nayak et~al.}(1999)\citenamefont{Nayak, Fisher, Ludwig,
  and Lin}}]{Nayak+99}
\bibinfo{author}{\bibfnamefont{C.}~\bibnamefont{Nayak}} \emph{et al.},
\bibinfo{journal}{Phys. Rev. B} \textbf{\bibinfo{volume}{59}},
  \bibinfo{pages}{15694} (\bibinfo{year}{1999}).

\bibitem[{\citenamefont{Rao and Sen}(2004)}]{Rao+04}
\bibinfo{author}{\bibfnamefont{S.}~\bibnamefont{Rao}} \bibnamefont{and}
  \bibinfo{author}{\bibfnamefont{D.}~\bibnamefont{Sen}},
  \bibinfo{journal}{Phys. Rev. B} \textbf{\bibinfo{volume}{70}},
  \bibinfo{pages}{195115} (\bibinfo{year}{2004}).

\bibitem{footnote3} An RG scenario similar to ours is also found in \cite{Yi+96,Nayak+99,Rao+04}.


\bibitem[{\citenamefont{Matveev et~al.}(1993)\citenamefont{Matveev, Yue, and
  Glazman}}]{Matveev+93}
\bibinfo{author}{\bibfnamefont{K.~A.} \bibnamefont{Matveev}} \emph{et al.},
\bibinfo{journal}{Phys. Rev. Lett.} \textbf{\bibinfo{volume}{71}},
  \bibinfo{pages}{3351} (\bibinfo{year}{1993}).


\end{thebibliography}

\end{document}